\documentclass[aps,pre,twocolumn,a4paper,superscriptaddress]{revtex4}%

\usepackage{epsfig,xcolor,soul,amsmath,amssymb,amsfonts,physics,graphicx}
\usepackage[utf8]{inputenc}

\usepackage{hyperref}
\hypersetup{
    colorlinks=true,
    linkcolor=blue,    
    urlcolor=blue,
    citecolor=blue
    }
    
\usepackage{lipsum}
\usepackage{bm}

\DeclareMathOperator{\Tra}{Tr}

\begin{document}

\title{Transverse vortices induced by modulated granular shear flows of elongated particles}

\author{Sára \surname{Lévay}}
\email{slevay@unav.es}
\affiliation{Institute for Solid State Physics and Optics, HUN-REN Wigner
Research Centre for Physics, Budapest, Hungary}
\affiliation{Departamento de Física y Matemática Aplicada, Facultad de Ciencias, Universidad de Navarra, Pamplona, Spain}
\author{Philippe \surname{Claudin}}
\affiliation{Physique et M\'ecanique des Milieux H\'et\'erog\`enes, PMMH UMR 7636 CNRS, ESPCI Paris, PSL Research University, Sorbonne Universit\'e, Universit\'e Paris Cit\'e, Paris, France}
\author{Ellák \surname{Somfai}}
\affiliation{Institute for Solid State Physics and Optics, HUN-REN Wigner Research Centre for Physics, Budapest, Hungary}
\author{Tamás \surname{Börzsönyi}}
\affiliation{Institute for Solid State Physics and Optics, HUN-REN Wigner Research Centre for Physics, Budapest, Hungary}

\begin{abstract}
We perform discrete element method (DEM) simulations of elongated grains in a shear cell for various particle aspect ratios and contact frictions, with an additional heterogeneous force perturbation in the flow direction. For a perturbation in the form of a single Fourier mode, we show that the response of the system consists of transverse secondary flows that average onto a pattern of four vortices. We also theoretically studied this phenomenon by generalizing the granular rheology $\mu(I)$ to the case of elongated grains and computing the linear response to such a perturbation. Even if the agreement between theory and simulations remains qualitative only, we can reproduce and understand the inversion of the vortex pattern when the cell aspect ratio is increased from a vertically to a horizontally elongated cell shape, emphasizing the key role of the second normal stress difference as well as the cell geometry.
\end{abstract} 

\date{\today}

\maketitle

\section{Introduction}
Secondary flows in granular systems have been observed in a variety of configurations. Vortex-like convective rolls have been studied in gravity-driven chute flows, where regions with very different densities develop, especially when highly agitated gas-like domains appear with considerably decreased density~\cite{forterrePRL2001, borzsonyiPRL2009, broduPRE2013, broduJFM2015}. Taylor vortex-like structures have also been seen in a granular flow~\cite{conwayN2004}, which are analogous to the well-known instability of liquid flows in the Taylor-Couette geometry~\cite{taylorPTRS1923}. The formation of heaps on the top surface of the granular material was also observed in a similar geometry~\cite{cabreraPRE2020}. In slower flows, where density variations are much smaller, density-driven secondary convection was studied in a cylindrical Couette device, as well as numerically in a split bottom shear cell~\cite{krishnarajNC2016, dsouzaEPJWC2017, dsouzaJFM2021}. Recently, a slow transverse flow was created by a bulldozing mechanism in driving granular material with a conveyor belt~\cite{escobar2024}.

The above-described observations were made with spherical or nearly spherical particles. For elongated grains, a peculiar secondary flow was observed in the cylindrical split bottom shear cell~\cite{wortelSM2015, fischerNJP2016}. The strength of this transverse flow was fill-height dependent, with the largest amplitude for intermediate fill heights, for which the shear zone has the largest extension. This secondary convection led to the formation of a considerable heap in the middle of the shear cell, which was not observed for spherical grains. Interestingly, reverse convection was recently observed for concave grains, leading to the formation of a dip in the middle of the cylindrical split bottom shear cell~\cite{mohammadiPRE2022}.

Secondary flows are also present in flows of granular suspensions~\cite{GuazzelliJFM2018}, which share part of their rheological behavior with dry granular systems~\cite{BoyerPRL2011, TrulssonPRL2012}. Secondary flows, possibly associated with migration of particles, are known to be induced by normal stress differences~\cite{RamachandranJFM2008}, as exemplified by the Weissenberg effect~\cite{yamamoto1958visco, oldroyd1958non, kundu1973normal, elson1982interaction} with viscoelastic fluids such as polymers. Normal stress differences in granular suspensions have been measured in experiments analyzing the curvature of the free surface of flows in a tilted trough~\cite{CouturierJFM2011}, or recording stress profiles in a torsional flow between discs~\cite{DboukJFM2013}, as well as in simulations~\cite{SetoJFM2018}. Interestingly, they get enhanced values for elongated particles such as fibers or dimers~\cite{SnookJFM2014, BounouaJOR2016, ButlerARFM2018, MariJOR2020}. The free-surface curvature of channelized gravity-driven flows of dry grains has also been associated with the presence of normal stress differences~\cite{McElwaineGM2012, GadalJFM2025}, especially in the context of the formation of levees~\cite{FelixEPSL2004, DeboeufPRL2006, RochaJFM2019}.

The flow of dense granular systems has been largely studied over the last two decades, in particular developing an empirical frictional rheology, known as the $\mu(I)$ rheology~\cite{GDRMiDiEPJE2004, daCruzPRE2005, jop2006constitutive, ForterreARFM2008, JopCRAS2015}. Starting from the idealized case of systems of mono-sized hard spheres, this rheology has been extended in various ways, for example, accounting for grain cohesion~\cite{RognonEPL2006, MandalPNAS2020, BlatnyJFM2024}, grain softness and contact friction~\cite{FavierPRF2017}, or active (self-propelled) grains~\cite{PeshkovEPL2016}. An important piece of work has also been done to properly accommodate, or regularize, this rheology in Navier-Stokes solvers~\cite{BarkerJFM2015, BarkerPRSA2017, GoddardPOF2018}. In case of the presence of strong gradients, a non-local rheology has been developed~\cite{PouliquenPTRSA2009, BouzidPRL2013, HenannPNAS2013}, for which an additional variable, the fluidity, has to be defined and modeled~\cite{BouzidEPJE2015, ZhangPRL2017, PoonPRE2023}. Finally, the kinetic theory for granular gases, pushed to the dense limit, recovers the features of the granular rheology and its non-local extensions~\cite{BerziPRF2024}.

For our present purpose, a key rheological aspect is to account for grain elongation, as e.g. with rods~\cite{borzsonyiPRL2012, guoPOF2013, berziSM2022}, multisphere particles~\cite{borzsonyiPRE2012, liuPRE2024, rahimPRF2024}, spherocylinders or ellipses~\cite{nagy2017rheology, TrulssonJFM2018, nagy2020flow}. These studies showed that, in a similar way to granular suspensions, these systems display enhanced normal stress differences. In our wish to understand the processes at work in the heaping problem, and to relate secondary transverse vortices to these normal stress differences, the idea is to investigate the flow of grains of arbitrary elongation aspect ratio $\alpha$ in the simpler configuration of the shear cell, where a non-homogeneous forcing is implemented in the flow direction. This forcing will be of arbitrarily small amplitude $\epsilon$ so that we can compute the response of the system at the linear order in this perturbation.

The structure of the paper is the following. In Sec.~\ref{Sec:gen_gran_rheology} we propose a generalized granular rheology in the case of elongated grains and study the case of a plane Couette flow. In Sec.~\ref{Sec:cons_pert} and \ref{Sec:linear_response} we analytically compute the linear response of a shear flow to a nonhomogeneous forcing in the flow direction. We introduce the discrete element method (DEM) and numerical setup to simulate such granular flows in Sec.~\ref{Sec:dem}. In Sec.~\ref{Sec:res} we present and discuss the results of these simulations in comparison to the theoretical expectations.

\section{Generalizing granular rheology}
\label{Sec:gen_gran_rheology}

\subsection{Reference isotropic case}
Assuming pressure isotropy, it has been proposed by Jop \textit{et al.}~\cite{jop2006constitutive} that the granular rheology, which relates the stress tensor $\sigma_{ij}$ to the velocity field $u_i$, takes the form
\begin{equation}
\sigma_{ij} = \Pi \left( \delta_{ij} + \mu \frac{\dot{\gamma}_{ij}}{\dot{\gamma}}\right),
\label{RheologyIsotropic}
\end{equation}
where $\Pi = \tfrac{1}{3} \sigma_{ii}$ is the pressure and $\dot{\gamma}_{ij} = \dot{\Gamma}_{ij} + \dot{\Gamma}_{ji}$ with $\dot{\Gamma}_{ij} = \partial_i u_j$ is the shear rate. The norm of the shear rate is here defined as $\dot{\gamma}^2 = \tfrac{1}{2} \dot{\gamma}_{ij} \dot{\gamma}_{ij}$. The effective friction $\mu$ is a function of the inertial number $I=\dot{\gamma}d/\sqrt{\Pi/\rho}$, where $d$ is the grain size and $\rho$ the grain mass density, and is empirically determined by measurements on experimental or numerical data. Although this rheology also comes with a second constitutive relation describing the behavior of the packing fraction $\phi(I)$, this expression (\ref{RheologyIsotropic}) must be consistent with the definition of $\Pi$, incompressibility: $\dot{\gamma}_{ii} = 0$. We will eliminate this possible issue by assuming later the quasistatic limit $I \to 0$.

\begin{figure*}
\centering
\includegraphics[width=0.85\textwidth]{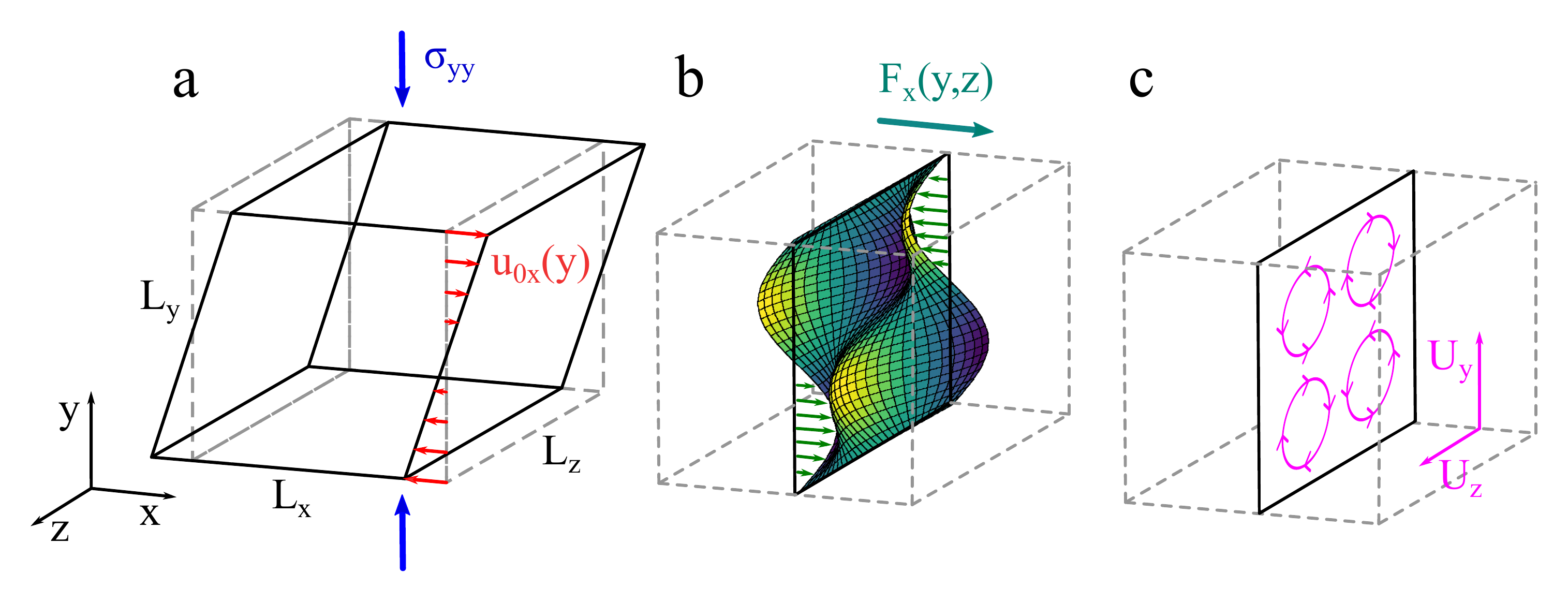}
\caption{Schematic drawing of the studied system. (a) The simulated cell has a dimension $L_x \times L_y \times L_z$. It is confined by $\sigma_{yy}$ in the vertical direction. Shear is applied in the $x$ direction, plus (b) an additional external force $F_{x}(y,z)$ in the $x$ direction (its magnitude and sign are shown by the surface and the green arrows). This results in the flow pattern shown in (c), four vertices in the velocity field in the $y$-$z$ plane.}
\label{Fig:system}
\end{figure*}

\subsection{Generalized constitutive relation}
We would like to generalize this rheology (\ref{RheologyIsotropic}) to the case of elongated grains, for which finite normal stress differences $N_1=\sigma_{xx} - \sigma_{yy}$ and $N_2 = \sigma_{yy} - \sigma_{zz}$ arise. To do so, one must notice that the following symmetries are required

(i) The normal stress differences depend on the direction of the velocity gradient and the flow, so they must change with the transformation $\dot{\bm{\Gamma}} \rightarrow \dot{\bm{\Gamma}}^T$. This means they must depend explicitly on $\dot{\Gamma}_{ij}$, as $\dot{\gamma}_{ij}$ is symmetric.

(ii) The normal stress differences are invariant to the reversal of the flow direction ($\dot{\Gamma}_{ij} \rightarrow -\dot{\Gamma}_{ij}$), so they can only depend on even order terms of the shear rate.

(iii) The new term should have a trace of zero, to keep the definition of $\Pi$ correct.

The simplest relation that fulfills these requirements is
\begin{widetext}
\begin{eqnarray}
\sigma_{ij} & = & \Pi \left[ \delta_{ij} + \mu \frac{\dot{\gamma}_{ij}}{\dot{\gamma}}
+ \nu_1 \frac{1}{\dot{\gamma}^2} \left( \dot{\Gamma}_{ki} \dot{\Gamma}_{kj} - \tfrac{1}{3} \dot{\Gamma}_{kl} \dot{\Gamma}_{kl} \delta_{ij} \right)
+ \nu_2 \frac{1}{\dot{\gamma}^2} \left( \dot{\Gamma}_{ik} \dot{\Gamma}_{jk} - \tfrac{1}{3} \dot{\Gamma}_{lk} \dot{\Gamma}_{lk} \delta_{ij} \right)
\right. \nonumber \\
& &+\, \left. \eta \frac{1}{\dot{\gamma}^2} \left( \dot{\Gamma}_{ik} \dot{\Gamma}_{kj} - \tfrac{1}{3} \dot{\Gamma}_{lk} \dot{\Gamma}_{kl} \delta_{ij} \right)
+ \eta \frac{1}{\dot{\gamma}^2} \left( \dot{\Gamma}_{ki} \dot{\Gamma}_{jk} - \tfrac{1}{3} \dot{\Gamma}_{kl} \dot{\Gamma}_{lk} \delta_{ij} \right)
\right],
\label{RheologyAnisotropic}
\end{eqnarray}
or, without index notation, in a coordinate-independent form
\begin{eqnarray}
\bm{\sigma} & = & \Pi \left[ \bm{1} + \mu \frac{\dot{\bm{\gamma}}}{\dot{\gamma}}
+ \nu_1 \frac{1}{\dot{\gamma}^2} \left( \dot{\bm{\Gamma}}^T\dot{\bm{\Gamma}} - \tfrac{1}{3} \Tra(\dot{\bm{\Gamma}}^T\dot{\bm{\Gamma}}) \bm{1} \right)
+ \nu_2 \frac{1}{\dot{\gamma}^2} \left( \dot{\bm{\Gamma}}\dot{\bm{\Gamma}}^T - \tfrac{1}{3} \Tra(\dot{\bm{\Gamma}}\dot{\bm{\Gamma}}^T) \bm{1} \right)
\right. \nonumber \\
& &+\, \left. \eta \frac{1}{\dot{\gamma}^2} \left( \dot{\bm{\Gamma}}\dot{\bm{\Gamma}} - \tfrac{1}{3} \Tra(\dot{\bm{\Gamma}}\dot{\bm{\Gamma}}) \bm{1} \right)
+ \eta \frac{1}{\dot{\gamma}^2} \left( \dot{\bm{\Gamma}}^T\dot{\bm{\Gamma}}^T - \tfrac{1}{3} \Tra(\dot{\bm{\Gamma}}^T\dot{\bm{\Gamma}}^T) \bm{1} \right)
\right],
\label{RheologyAnisotropicNoindex}
\end{eqnarray}
\end{widetext}
with $\dot{\gamma}^2 = \tfrac{1}{2} \Tra(\dot{\bm{\gamma}} \dot{\bm{\gamma}}^T)$. The last two terms must have the same factor $\eta$ to ensure that the stress tensor is symmetric. The coefficients $\mu$, $\nu_1$, $\nu_2$, and $\eta$ depend on the inertial number $I$, but also on microscopic parameters such as the grain aspect ratio $\alpha$ or the grain contact friction $\mu_p$ \cite{nagy2017rheology, nagy2020flow, SrivastavaJFM2021}. This rheological form is similar to a Rivlin-Ericksen stress closure, as proposed for granular flows in \cite{McElwaineGM2012, SrivastavaJFM2021, KimFP2023, GadalJFM2025}.

\subsection{Plane Couette flow}
The granular rheology is usually studied and calibrated in the geometry of a plane Couette flow. In this case, we consider a cell sheared in the $x$ direction with the gradient in the $y$ ("vertical") direction, and $z$ being the neutral ("transverse") direction. This cell has dimensions $L_x$, $L_y$, and $L_z$ along those three axes, and we take the origin of the coordinates at the center of the box. The grains in this cell are subjected to a given pressure $\Pi_0$. The flow in the homogeneous case, which will be the base state of the calculation later on, is characterized by a linear velocity profile, associated with a reference shear rate $\dot{\gamma}_0$:
\begin{eqnarray}
\bm u_0 = &&
\begin{bmatrix}
\dot{\gamma}_0 y\\0\\0
\end{bmatrix},
\label{BaseStateVelocityDotGamma1} \\
\dot{\bm{\Gamma}}_0 = &&
\begin{bmatrix}
0&0&0\\
\dot{\gamma}_0&0&0\\
0&0&0
\end{bmatrix},
\label{BaseStateVelocityDotGamma2} \\
\dot{\bm{\gamma}}_0 = && \dot{\bm{\Gamma}}_0 + \dot{\bm{\Gamma}}_0^T =
\begin{bmatrix}
0&\dot{\gamma}_0&0\\
\dot{\gamma}_0&0&0\\
0&0&0
\end{bmatrix}.
\label{BaseStateVelocityDotGamma3}
\end{eqnarray}
In this configuration for which only $\dot{\Gamma}_{yx} \neq 0$, both $\dot{\bm{\Gamma}}_0^2$ and $(\dot{\bm{\Gamma}}_0^T)^2$ are zero, so that the second line of Eq.~(\ref{RheologyAnisotropicNoindex}) disappears and the coefficient $\eta$ cannot be measured in such a configuration. To measure it, a configuration with extensional or compressional flows is required. It will then be left adjustable in the following. In contrast, $\nu_1$ and $\nu_2$ are related to the calibrated first and second normal stress differences as
\begin{equation}
\nu_1 = \frac{N_1 + N_2}{\Pi} , \qquad \nu_2 = \frac{N_2}{\Pi},
\label{nu1nu2N1N2}
\end{equation}
whose values are taken from Nagy \textit{et al.}~\cite{nagy2020flow} (see also Fig.~\ref{Fig:theory} in Appendix~\ref{App:th}). The traces are related as
\begin{eqnarray}
\dot{\gamma}_0^2 & = & \tfrac{1}{2} \Tra(\dot{\bm{\gamma}}_0 \dot{\bm{\gamma}}_0^T) = \tfrac{1}{2}
\Tra(\dot{\bm{\Gamma}}_0^2 + \dot{\bm{\Gamma}}_0 \dot{\bm{\Gamma}}_0^T + \dot{\bm{\Gamma}}_0^T \dot{\bm{\Gamma}}_0
+ (\dot{\bm{\Gamma}}_0^T)^2)
\nonumber \\
& = & \Tra(\dot{\bm{\Gamma}}_0^2 + \dot{\bm{\Gamma}}_0 \dot{\bm{\Gamma}}_0^T) = \Tra(\dot{\bm{\Gamma}}_0 \dot{\bm{\Gamma}}_0^T).
\label{normgammadot}
\end{eqnarray}

As mentioned above, we will restrict to the quasistatic limit for which $I_0=\dot{\gamma}_0d/\sqrt{\Pi_0/\rho} \ll 1$. Under this assumption, the stress tensor in this base state finally reads
\begin{equation}
\bm{\sigma}_0 = \Pi_0 \left( \pi_{c} \bm{1} + 
\begin{bmatrix}
\nu_{1c}&\mu_{c}&0\\
\mu_{c}&\nu_{2c}&0\\
0&0&0
\end{bmatrix}
\right) = \Pi_0 \left( \pi_{c} \bm{1} + \bm{\Sigma}_0 \right),
\label{BaseStateStressTensor}
\end{equation}
where $\pi_{c} = 1 - \tfrac{1}{3}(\nu_{1c} + \nu_{2c})$, and where the subscript "c" indicate that $\mu$, $\nu_1$ and $\nu_2$ are taken constants (independent of the shear rate, but still functions of $\alpha$ and $\mu_p$, see Appendix~\ref{App:th}).

\section{Conservation equations and perturbation}
\label{Sec:cons_pert}

The equations that govern the dynamics of the system are the conservation of mass and momentum. Assuming incompressibility ($\phi \sim \phi_c$ is constant), these equations read
\begin{eqnarray}
\partial_j u_j & = & 0,
\label{MassCons} \\
\rho\phi_c u_j \partial_j u_i & = & \partial_j \sigma_{ij} + b_i.
\label{MomentumCons}
\end{eqnarray}
Here, $b_i$ is an external body force density. It is an arbitrary perturbation. We take it in the following form:
\begin{equation}
\bm b =
\epsilon \frac{\Pi_0}{d}
\begin{bmatrix}
f(y,z)\\0\\0
\end{bmatrix},
\label{eff}
\end{equation}
where $\epsilon$ is its vanishing dimensionless amplitude and $f$ its normalized profile in the $y$-$z$ plane. We take no component along the $y$ and $z$ directions.

Assuming that $\epsilon \ll 1$, the response of velocity and pressure to this perturbation can be written in linear order with respect to the base state as
\begin{eqnarray}\label{Eq:response}
\bm u = && \bm{u}_0 + \epsilon \dot{\gamma}_0 d 
\begin{bmatrix}
U_x(y, z)\\
U_y(y, z)\\
U_z(y, z)
\end{bmatrix} + o(\epsilon^2), \\
\Pi = && \Pi_0 \left( 1 + \epsilon P(y, z) \right) + o(\epsilon^2),
\end{eqnarray}
where $U_i$ and $P$ are unknown functions that we wish to relate to $f$.
With these expressions, one can compute the shear rate
\begin{equation}
\dot{\bm{\Gamma}} = \dot{\gamma}_0 
\begin{bmatrix}
0&0&0\\
1 + \epsilon d \partial_y U_x&\epsilon d \partial_y U_y&\epsilon d \partial_y U_z\\
\epsilon d \partial_z U_x&\epsilon d \partial_z U_y&\epsilon d \partial_z U_z
\end{bmatrix} + o(\epsilon^2),
\end{equation}
\begin{widetext}
\noindent
the symmetric shear rate
\begin{equation}
\dot{\bm{\gamma}} = \dot{\gamma}_0 
\begin{bmatrix}
0&1 + \epsilon d \partial_y U_x&\epsilon d \partial_z U_x\\
1 + \epsilon d \partial_y U_x&2\epsilon d \partial_y U_y&\epsilon d (\partial_z U_y + \partial_y U_z)\\
\epsilon d \partial_z U_x&\epsilon d (\partial_z U_y + \partial_y U_z)&2\epsilon d \partial_z U_z
\end{bmatrix} + o(\epsilon^2),
\end{equation}
and the second-order terms
\begin{eqnarray}
\dot{\bm{\Gamma}}^T \dot{\bm{\Gamma}} = \dot{\gamma}_0^2
\begin{bmatrix}
1 + 2 \epsilon d \partial_y U_x&\epsilon d \partial_y U_y&\epsilon d \partial_y U_z\\
\epsilon d \partial_y U_y&0&0\\
\epsilon d \partial_y U_z&0&0
\end{bmatrix}
+ o(\epsilon^2) , & \qquad &
\dot{\bm{\Gamma}} \dot{\bm{\Gamma}}^T = \dot{\gamma}_0^2
\begin{bmatrix}
0&0&0\\
0&1 + 2 \epsilon d \partial_y U_x&\epsilon d \partial_z U_x\\
0&\epsilon d \partial_z U_x&0
\end{bmatrix}
+ o(\epsilon^2) ,
\label{SecondOrderTerms1} \\
\dot{\bm{\Gamma}} \dot{\bm{\Gamma}} = \dot{\gamma}_0^2
\begin{bmatrix}
0&0&0\\
\epsilon d \partial_y U_y&0&0\\
\epsilon d \partial_z U_y&0&0
\end{bmatrix}
+ o(\epsilon^2) , & \qquad &
\dot{\bm{\Gamma}}^T \dot{\bm{\Gamma}}^T = \dot{\gamma}_0^2
\begin{bmatrix}
0&\epsilon d \partial_y U_y&\epsilon d \partial_z U_y\\
0&0&0\\
0&0&0
\end{bmatrix}
+ o(\epsilon^2) .
\label{SecondOrderTerms2} 
\end{eqnarray}
Even if $\dot{\bm{\Gamma}}^2$ and $(\dot{\bm{\Gamma}}^T)^2$ now do not vanish, their traces are still zero to the second order in $\epsilon$, so that
\begin{equation}
\dot{\gamma} = \sqrt{\tfrac{1}{2} \Tra(\dot{\bm{\gamma}} \dot{\bm{\gamma}}^T)} = \sqrt{\Tra(\dot{\bm{\Gamma}} \dot{\bm{\Gamma}}^T)} =  \dot{\gamma}_0 (1 + \epsilon d \partial_y U_x) + o(\epsilon^2).
\end{equation}
Finally, the stress tensor is
\begin{equation}
\bm \sigma = \Pi \left( \pi_{c} \bm{1} + \bm{\Sigma} \right) = \Pi_0 \left( 1 + \epsilon P(y, z)
\right) \left( \pi_{c} \bm{1} + \bm{\Sigma}_0 + \epsilon \bm S \right) = \bm{\sigma}_0 + \epsilon
\left( P \bm{\sigma}_0 + \Pi_0 \bm S \right) + o(\epsilon^2),
\end{equation}
where, using Eq.~\eqref{RheologyAnisotropicNoindex} and the above expressions, we obtain
\begin{equation}
\bm S = d
\begin{bmatrix}
0&(\nu_{1c}+\eta_{c}) \partial_y U_y&\mu_{c} \partial_z U_x + \nu_{1c} \partial_y
U_z + \eta_{c} \partial_z U_y\\
(\nu_{1c}+\eta_{c}) \partial_y U_y&2 \mu_{c} \partial_y U_y&\mu_{c} (\partial_z
U_y + \partial_y U_z) + \nu_{2c} \partial_z U_x\\
\mu_{c} \partial_z U_x + \nu_{1c} \partial_y U_z + \eta_{c} \partial_z U_y&\mu_{c} (\partial_z U_y +
\partial_y U_z) + \nu_{2c} \partial_z U_x &2 \mu_{c} \partial_z U_z
\end{bmatrix}.
\end{equation}
\end{widetext}

\begin{figure*}
\centering
\includegraphics[width=0.8\textwidth]{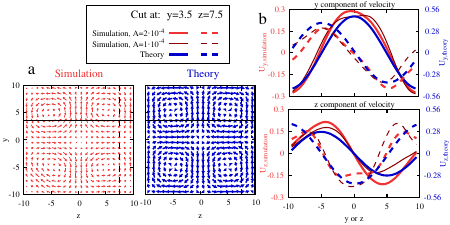}
\caption{Comparison of the DEM simulations (red) and calculations based on the generalized granular rheology (blue) for $\alpha{=}2$, $\mu_p{=}0.1$, and $L_z/L_y{=}1$ [presented also in Fig.~\ref{Fig:cellAR}(b)]. (a) The vortex pattern of the $y$-$z$ plane obtained by the coarse-grained averaging of the velocities in simulations, and the expected pattern according to the theory. The width of the arrows represents the magnitude of velocities. (b) Cross sections of the velocity profiles at $y=3.5$ (continuous lines) and $z=7.5$ (dashed lines). In the case of simulations, we present two cross sections obtained by applying different perturbation strengths. Stronger perturbation corresponds to the light red (thicker lines, $A=2\times10^{-4}$), while weaker perturbation is represented by the dark red (narrower lines, $A=1\times10^{-4}$).
}
\label{Fig:velprofs}
\end{figure*}

\section{Computation of the linear response}
\label{Sec:linear_response}

\subsection{Governing equations}
Taking linear terms in $\epsilon$ in Eqs. (\ref{MassCons}) and (\ref{MomentumCons}), we have
\begin{eqnarray}
\partial_i U_i = 0 ,
\end{eqnarray}
and
\begin{eqnarray}
\rho \phi_{c} \dot{\gamma}_0 d \left( U_j \partial_j u_{0,i} + u_{0,j} \partial_j U_i \right) \nonumber \\
= \sigma_{0,ij} \partial_j P + \Pi_0 \partial_j S_{ij}
+ \frac{\Pi_0}{d} \begin{bmatrix} f\\0\\0 \end{bmatrix}.
\end{eqnarray}
Substituting the previous expressions and using the fact that $\partial_x (.) = 0$, we obtain
\begin{eqnarray}
\partial_y U_y + \partial_z U_z &=& 0 , \label{divuxhomogeneous}\\
\frac{\rho \phi_{c} \dot{\gamma}_0^2 d}{\Pi_0} U_y &=& \mu_c \left( \partial_y P + d \partial_{zz} U_x \right) \nonumber \\
& & +\, \nu_{1c} d (\partial_{zy} U_z + \partial_{yy} U_y) \\
& & +\, \eta_{c} d (\partial_{yy} U_y + \partial_{zz} U_y) + \frac{1}{d} f, \nonumber\\
0 &=& \left( \pi_{c} + \nu_{2c} \right) \partial_y P + 2 \mu_{c} d \partial_{yy} U_y \nonumber \\ 
& & +\, \mu_{c} d (\partial_{zz} U_y + \partial_{zy} U_z) \\
& & +\, \nu_{2c} d \partial_{zz} U_x , \nonumber \\
0 &=& \pi_{c} \partial_z P + \mu_{c} d (\partial_{yz} U_y + \partial_{yy}U_z) \nonumber \\
& & +\, \nu_{2c} d \partial_{yz} U_x + 2 \mu_{c} d \partial_{zz} U_z,
\end{eqnarray}
which, once rearranged and simplified (using in particular $\partial_y U_y + \partial_z U_z = 0$), gives
\begin{eqnarray}
\partial_y U_y + \partial_z U_z &=& 0 ,\\
\frac{\rho \phi_{c} \dot{\gamma}_0^2 d^2}{\Pi_0} U_y &=& \mu_c d \left( \partial_y P + d \partial_{zz} U_x \right) \nonumber \\
& & +\, \eta_{c} d^2 (\partial_{yy} U_y + \partial_{zz} U_y) + f, \qquad \\
- \mu_{c} d (\partial_{zz} U_y + \partial_{yy} U_y) &=& \left( \pi_{c} + \nu_{2c} \right) \partial_y P \nonumber \\ 
& & +\, \nu_{2c} d \partial_{zz} U_x ,\\
- \mu_{c} d (\partial_{zz} U_z + \partial_{yy} U_z) &=& \pi_{c} \partial_z P + \nu_{2c} d \partial_{yz} U_x .
\end{eqnarray}
Because we work in the limit of a vanishing inertial number, the LHS term in the second equation $\phi_c I_0^2 U_y$ can be neglected, so that
\begin{equation}
\mu_c d \left( \partial_y P + d \partial_{zz} U_x \right) + \eta_{c} d^2 (\partial_{yy} U_y + \partial_{zz} U_y) + f \approx 0.
\end{equation}
With this, we finally obtain the following governing linear equations
\begin{eqnarray}
\partial_y U_y + \partial_z U_z = 0, \,\,\, &&
\label{LinEq1} \\
\partial_y P + d\partial_{zz} U_x + \frac{\eta_{c} d}{\mu_c} (\partial_{yy} U_y + \partial_{zz} U_y) + \frac{1}{\mu_c d} f = 0, \,\,\, &&
\label{LinEq2} \\
- \! \left( \! \mu_{c} - \frac{\nu_{2c} \eta_c}{\mu_c} \! \right) d (\partial_{zz} U_y + \partial_{yy} U_y) = \pi_{c} \partial_y P - \frac{\nu_{2c}}{\mu_c d} f, \,\,\, &&
\label{LinEq3} \\
- \mu_{c}d (\partial_{zz} U_z + \partial_{yy} U_z) = \pi_{c} \partial_z P + \nu_{2c}d \partial_{yz} U_x. \,\,\, &&
\label{LinEq4}
\end{eqnarray}
To recall, the first of these four equations comes from mass conservation assuming incompressibility Eq.~(\ref{MassCons}) and the next three express momentum conservation Eq.~(\ref{MomentumCons}).

\subsection{Single-mode solution}
Simple modes of the above Eqs. (\ref{LinEq1}) to (\ref{LinEq4}) are the form of the product of sine and cosine functions:
\begin{eqnarray}
U_x & = & a_x \sin(k_y y/d)\cos (k_z z/d), 
\label{Uxsimple} \\
U_y & = & a_y \sin(k_y y/d)\cos (k_z z/d),
\label{Uysimple} \\
U_z & = & a_z \cos(k_y y/d)\sin (k_z z/d),
\label{Uzsimple} \\
P & = & a_p \cos(k_y y/d)\cos (k_z z/d),
\label{Psimple} 
\end{eqnarray}
with dimensionless wavenumbers $k_y=2\pi d/L_y$ and $k_z=2\pi d/L_z$. Assuming the perturbation itself has a profile along a mode
\begin{equation}
f(y,z) = \sin(k_y y/d)\cos (k_z z/d),
\end{equation}
then Eqs. (\ref{LinEq1}) to (\ref{LinEq4}) simplify into
\begin{eqnarray}
k_y a_y + k_z a_z & = & 0,\\
k_z^2 a_x + \frac{\eta_c}{\mu_c} (k_y^2+k_z^2) a_y + k_y a_p & = & \frac{1}{\mu_c},\\
\left( \mu_c - \frac{\nu_{2c} \eta_c}{\mu_c} \right) (k_y^2+k_z^2) a_y + \pi_c k_y a_p & = & - \frac{\nu_{2c}}{\mu_c},\\
\nu_{2c} k_y k_z a_x + \mu_c (k_y^2+k_z^2) a_z + \pi_c k_z a_p & = & 0,
\end{eqnarray}
\begin{widetext}
\noindent
whose solution is
\begin{eqnarray}\label{Eq:finalx}
a_x & = & \frac{1}{\mu_c k_z^2} \times \frac{\pi_c + \left( \frac{k_y}{k_z} \right)^2 (\pi_c + \nu_{2c})}{\pi_c (1-\beta_c) + \left( \frac{k_y}{k_z} \right)^2 \left[ \pi_c (1+\beta_c) - \nu_{2c} (1-\beta_c) \right]},\\
\label{Eq:finaly}
a_y & = & \frac{\nu_{2c}}{\mu_c^2 (k_y^2 + k_z^2)} \times \frac{-\pi_c + \left( \frac{k_y}{k_z} \right)^2 (\pi_c + \nu_{2c})}{\pi_c (1-\beta_c) + \left( \frac{k_y}{k_z} \right)^2 \left[ \pi_c (1+\beta_c) - \nu_{2c} (1-\beta_c) \right]},\\
\label{Eq:finalz}
a_z & = & - \frac{\nu_{2c}}{\mu_c^2 (k_y^2 + k_z^2)} \left( \frac{k_y}{k_z} \right) \times \frac{-\pi_c + \left( \frac{k_y}{k_z} \right)^2 (\pi_c + \nu_{2c})}{\pi_c (1-\beta_c) + \left( \frac{k_y}{k_z} \right)^2 \left[ \pi_c (1+\beta_c) - \nu_{2c} (1-\beta_c) \right]},\\
a_p & = & - \frac{2 \nu_{2c}}{\mu_c k_z^2} \times \frac{k_y}{\pi_c (1-\beta_c) + \left( \frac{k_y}{k_z} \right)^2 \left[ \pi_c (1+\beta_c) - \nu_{2c} (1-\beta_c) \right]},\label{Eq:finalp}
\end{eqnarray}
where we define $\beta_c = \eta_c \nu_{2c}/\mu_c^2$ for short. Note that $a_y$ and $a_z$ are simply proportional, as expected from a velocity field of vanishing divergence, but homogeneous in the $x$ direction Eq.~(\ref{divuxhomogeneous}): $a_z = - a_y \, k_y/k_z$.
\end{widetext}

As for boundary conditions, the above expressions satisfy
\begin{eqnarray}
U_y (y = \pm L_y/2) & = & 0, \label{BCy}\\
U_z (z = \pm L_z/2) & = & 0, \label{BCz}
\end{eqnarray}
as they should, so that there is no flux through the sides of the cell. Furthermore, the constant $-a_p$ can be added to $P$ to make it vanish at the cell corners. This solution shows a velocity field with four vortices in the $y$-$z$ plane, as displayed in Figs. \ref{Fig:system}(c) and \ref{Fig:velprofs}(a).

The above solution can be generalized for higher wavenumbers $n_y \times 2\pi d/L_y$ and $n_z \times 2\pi d/L_z$, where $n_y$ and $n_z$ are integers, in association with the Fourier decomposition of an arbitrary forcing profile $f(y,z)$.

\section{DEM simulation}
\label{Sec:dem}

\subsection{Numerical setup}
To study the behavior of a sheared granular system in the presence of spatially modulated additional body force we performed DEM simulations in the three-dimensional plane Couette geometry (see Fig.~\ref{Fig:system}), with periodic boundary conditions in the $x$ (flow) and $z$ (neutral) directions and Lees-Edwards boundary conditions in the $y$ (velocity gradient) direction. This way wall effects could be eliminated. The particles were frictional spherocylinders with length-to-diameter aspect ratio $\alpha=\ell/d$. In addition, to avoid crystallization, some amount of size polydispersity was introduced: while we kept the aspect ratio constant, we drew the diameters of the particles from a uniform distribution with a ratio of standard deviation to mean of $10\%$. The length, time, and mass units of the simulation were set implicitly by setting the mean particle diameter $d$, density $\rho$, and contact stiffness $k$ (equal for the normal and tangential force) to unity.

We created the initial conditions of random particle orientation with overdamped dynamics, and afterward sheared the system at constant shear rate $\dot{\gamma}$. During shear, we employed stress control, where one side of the box, $L_y$, was adjusted by a feedback loop such that the corresponding normal stress $\sigma_{yy}$ fluctuated around a specified value $-p_y = 10^{-3}$. We kept $L_x$ and $L_z$ fixed. The simulations were run with an inertial number on the order of $I \simeq 10^{-2}$, which is a good compromise between a shear rate small enough for the system to behave in the quasistatic regime, but large enough to achieve significant deformation in a reasonable CPU time. Recall that deformations as large as $\gamma \simeq 200$ are required to see secondary flows that average out from the noise. For more details about the DEM simulations, such as the interaction forces, we refer the reader to \cite{nagy2017rheology,nagy2020flow}.

To implement the external body perturbation [term $b_i$ in RHS of Eq.~(\ref{MomentumCons})], we applied a position-dependent force at the center of every particle, acting in the $x$ direction:
\begin{equation}
F_x(y,z) = A \sin(k_y y/d)\cos (k_z z/d).
\label{FxinSimulations}
\end{equation}
The constant $A$ is typically set to a small value in the order of $10^{-4}$ so that it acts as a perturbation with respect to the main shear flow. It relates to $\epsilon$ introduced in the previous section as $A=\epsilon \frac{\Pi_0}{d} V_P/\phi_c$, where $V_P$ is the average volume of the particles. Importantly, as discussed in detail in the next section, this small value for $A$ already induces significant nonlinearities for the streamwise response, although lower values would make the measurement of the transverse response (the vortices we are interested in) very difficult given the cell deformation we can reach ($\gamma \simeq 200$) and the noise associated with grain collisions.

We ran several DEM simulations to measure the response of the system and compare it to the theoretical predictions. We varied the particle aspect ratio ($\alpha$), the particle-particle friction coefficient ($\mu_p$), the strength of the perturbation ($A$), and the cell aspect ratio ($L_z/L_y$). For each case, we ran five independent realizations and averaged over them. All measurements were taken in the stationary state after a deformation of at least $\gamma=10$. All quantities of interest were time-averaged for a duration corresponding to a deformation of at least $\gamma=100$. The number of particles in the cell was $4000$, except for the largest cell aspect ratio for which it was increased to $10300$.

\subsection{Calculation of the response}
To obtain the response of the system to the force perturbation, we calculated the coarse-grained (CG)~\cite{goldhirsch2010stress} velocity profiles in the $y$-$z$ plane using the particle length $\ell$ as the CG scale and also averaging in the $x$ direction along which the system is homogeneous. This resulted in the velocity vector $\bm{u}$ [see Eq.~(\ref{Eq:response})]. After subtracting the base velocity $\bm{u_0}$ and dividing by $\epsilon \dot{\gamma}_0 d$, we obtained the rescaled velocity response $U_i(y,z)$ (here for $i=y,z$ only), which have been fitted according to Eqs. (\ref{Uxsimple}) to (\ref{Uzsimple}). This way we obtain the values of the coefficients $a_i$, which can be directly compared to those predicted by the theory in Eqs. (\ref{Eq:finalx}) to (\ref{Eq:finalp}). As said above, to compute the value of $\epsilon$ from that of $A$, one needs the main pressure $\Pi_0$, which was calculated from the values of $N_1$, $N_2$ (taken from Nagy et al~\cite{nagy2020flow} at the value of $I=10^{-2}$, see also Fig.~\ref{Fig:theory} in Appendix~\ref{App:th}), and the imposed normal stress $\sigma_{yy}$. As $N_1$ and $N_2$ depend on $\alpha$ and $\mu_p$, the value of $\Pi_0$ varied from case to case but was typically in the order of $\Pi_0\sim10^{-3}$, i.e., expectedly comparable to $\sigma_{yy}$.

\begin{figure*}[t!]
\centering
\includegraphics[width=0.9\textwidth]{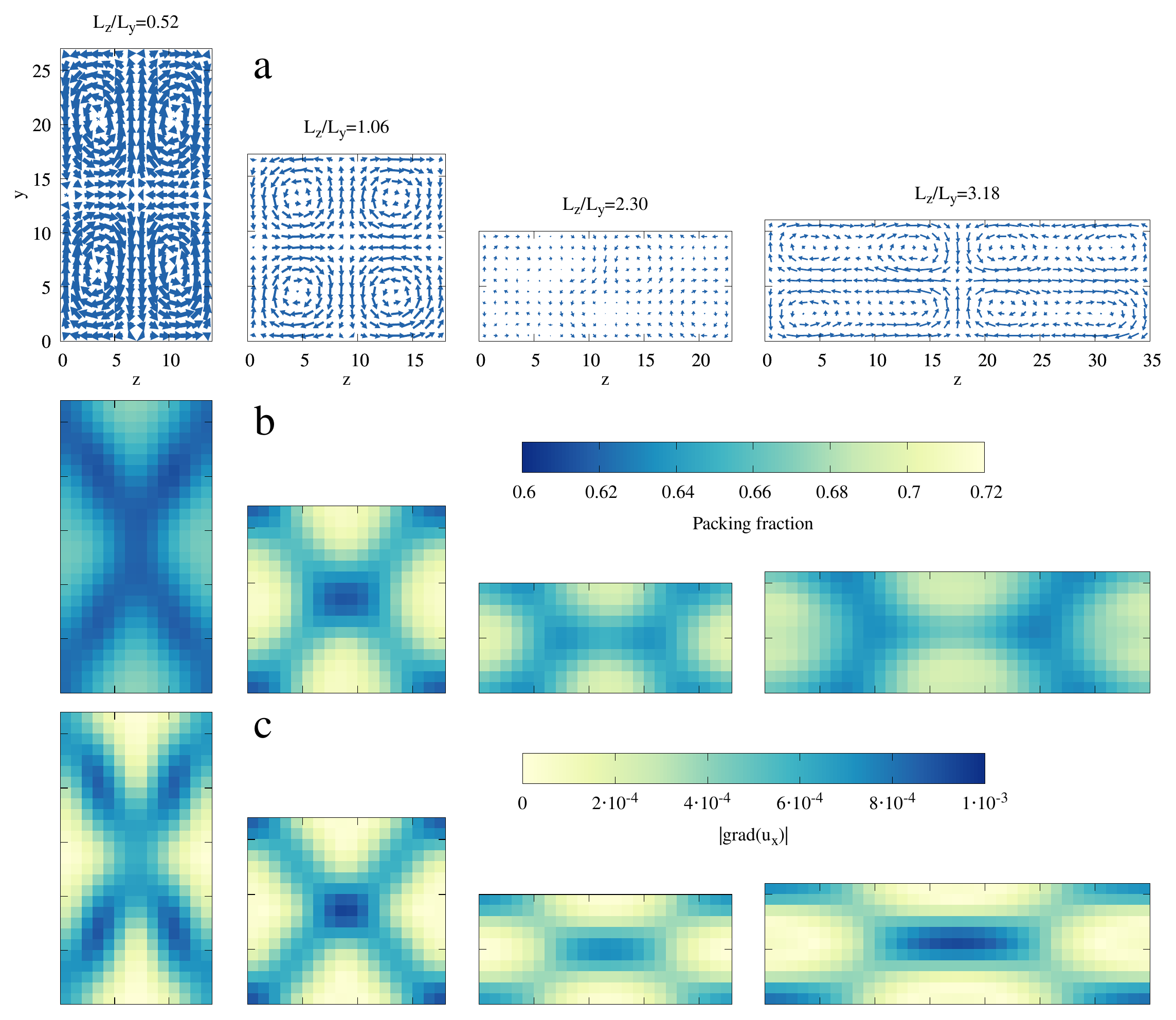}
\caption{(a) Coarse-grained velocity profiles of cells with different cell aspect ratios obtained by the DEM simulations for $\alpha=1.5$, $\mu_p=0.01$, and $A=2\times10^{-4}$. The width of the arrows represents the magnitude of velocities. For the time evolution of the vortex pattern, we refer the reader to the SM videos~\cite{SM}. (b) Corresponding packing fraction and (c) velocity gradient $|\text{grad} (u_x)|$ maps. Note that for the largest cell aspect ratio $L_z/L_y=3.18$, the particle number was increased to $10300$, while in all the other cases it is $4000$. Note also that the colorbars of panels (b) and (c) are reversed to emphasize the similarity of the patterns.}
\label{Fig:cellAR_profiles}
\end{figure*}

\section{Results and discussion}
\label{Sec:res}

First, we present cross sections of the velocity field in the $y$-$z$ plane, which show the secondary circulation resulting from the modulated forcing in the $x$ direction in a sheared system (as sketched in Fig.~\ref{Fig:system}). In Fig.~\ref{Fig:velprofs} we show the results of simulations and theory for a cell with aspect ratio $L_z/L_y=1$, filled with particles of aspect ratio $\alpha=2$ and friction coefficient $\mu_p=0.1$. The overall agreement between the numerics (red) and the calculations (blue) made according to the generalized granular rheology is good; they both show two pairs of counterrotating vortices. In Fig.~\ref{Fig:velprofs}(a) we show the CG velocity field in the $y$-$z$ plane,  while Fig.~\ref{Fig:velprofs}(b) compares cross sections of the velocity field at $y=3.5$ (continuous line) and $z=7.5$ (dashed line), for the $y$ and $z$ components of velocity, and one can see the sinusoidal shape in both cases, as anticipated by Eqs. (\ref{Uysimple}) and (\ref{Uzsimple}).

We tested different perturbation strengths $A=1\times10^{-4}$, $2\times10^{-4}$, $3\times10^{-4}$, and $5\times10^{-4}$. This perturbation has to be strong enough to allow for a measure of the response out of the noise, but ideally not too large to be compatible with the linear analysis developed above. Such values $A \simeq 10^{-4}$ correspond to $\epsilon$ on the order of a few percents, and this is indeed the relative magnitude of the transverse components of the velocity response. However, for the same $A$, that of the stream-wise velocity is significantly larger, on the order of a few tens of percents. This can actually be understood from the linear theory: comparing the response coefficients $a_x$ and $a_y$ in Eqs.~(\ref{Eq:finalx}) and (\ref{Eq:finaly}) respectively, one can see that $a_x$ must be at least larger than $a_y$ by a factor of $\mu_c/\nu_{2c} \simeq 10$. In addition, as the numerator of $a_x$ comes as a sum, whereas that of $a_y$ comes as a difference, this provides an additional factor of a few units. As a consequence, the response to the perturbation in the $x$ direction is significantly larger than that in the transverse direction. This inevitably puts us away from a proper linear regime as we cannot reliably measure the transverse response with $A$ smaller than $10^{-4}$ (too much noise for reasonable time averaging, a deformation $\gamma \simeq 200$). However, we observe that, for the two smallest values of the perturbation we have used ($A=1\times10^{-4}$ and $2\times10^{-4}$), the transverse velocity response is linear in the sense that, not only the profile shape is sinusoidal, but the curves collapse once rescaled by $A$, as shown in Fig.~\ref{Fig:velprofs}(b). For the larger values of $A$ we have tested, the strength of the response can not be rescaled anymore.

\begin{figure*}[t]
\centering
\includegraphics[width=0.85\textwidth]{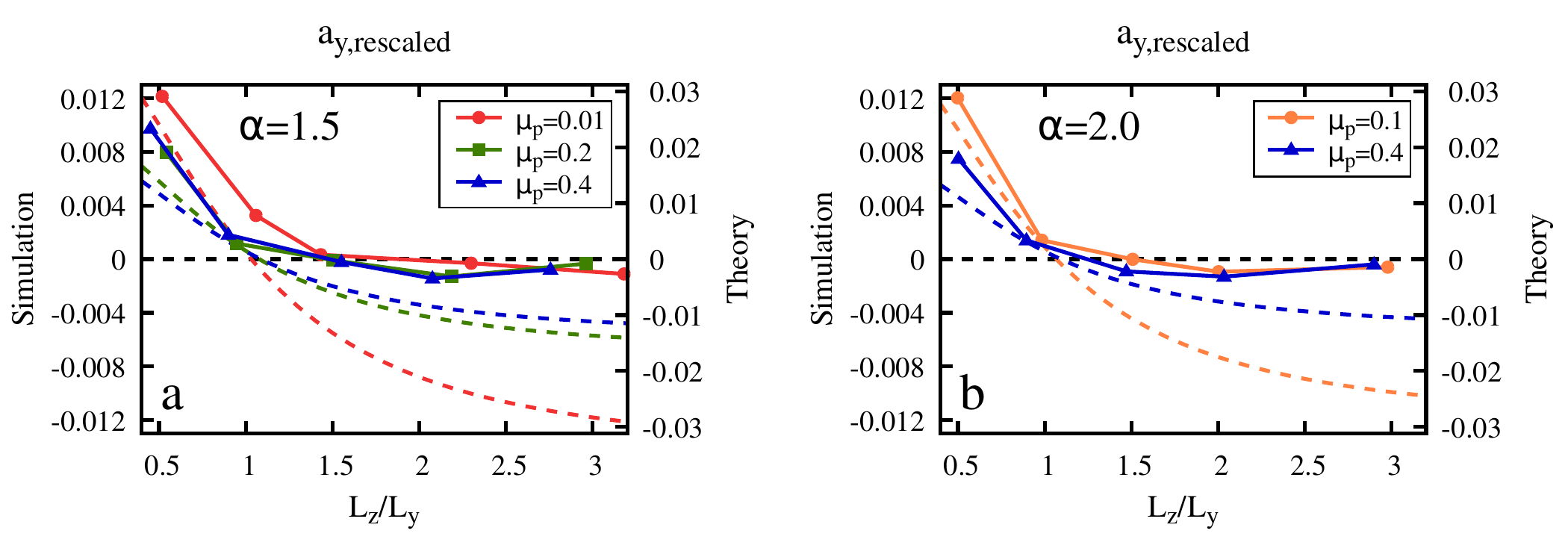}
\caption{Response of the system, rescaled by the system size [$a_{y\text{,rescaled}}{=}a_y(L_y^2{+}L_z^2)/(L_y^2 L_z^2)$], as a function of cell aspect ratio for two particle aspect ratios (a) $\alpha{=}1.5$ and (b) $\alpha{=}2$, and for different particle coefficients of friction $\mu_p$ (legends). The strength of perturbation is $A=2\times10^{-4}$. Dashed lines show the expected values according to the theory. The symbols show the results of simulations, averaged over five different realizations. Their span is in the order of the point size. Importantly, note the different vertical axes on these graphs for simulation results (left) and theoretical predictions (right) --- a qualitative-only agreement is reached.}
\label{Fig:cellAR}
\end{figure*}

A strong and robust prediction of the theory is that the sense of rotation of the vortices should change when the cell aspect ratio  $L_z/L_y$ increases. Indeed, our numerical simulations show this inversion as we see in the cross sections of the velocity fields presented in Fig.~\ref{Fig:cellAR_profiles}(a), or in Fig.~\ref{Fig:cellAR} where we see the change of sign of the vortex amplitude as a function of $L_z/L_y$. The origin of this inversion in the theoretical description can be seen in the expressions for the amplitudes $a_y$ and $a_z$ in Eqs. (\ref{Eq:finaly}) and (\ref{Eq:finalz}), where the numerator changes sign for $k_y/k_z = L_z/L_y = \sqrt{\pi_c/(\pi_c + \nu_{2c})} \gtrsim 1$, while the denominator stays positive.

Our numerical simulations also allow us to extract information about the spatial distribution of the packing fraction in the system. In Figs. \ref{Fig:cellAR_profiles}(b) and \ref{Fig:cellAR_profiles}(c), we show the maps for the packing fraction as well as for the norm of the gradient of the $x$ component of the velocity $|\text{grad} (u_x)|$ corresponding to the systems presented in Figs.~\ref{Fig:cellAR_profiles}(a). As we see, there is a correlation between shear rate and packing fraction, which can be related to two effects (i) In our system, the inertial number $I$ reaches a value of $10^{-2}$, meaning that inertial effects are already present, although with small amplitude. Such effects lead to slightly decreased density due to more intensive collisions at the locations with larger shear rates. (ii) Spatial variation of the direction of the shear gradient will lead to a spatial modulation of the shear-induced alignment of the grains, which can also lead to decreased packing fraction if it causes frustration in the grain orientation. Irrespective of the cause of the spatial variation of the packing fraction, the most important aspect is that we see that for the secondary flow, the change in the rotation direction is not accompanied by a change in the packing fraction map.

In Fig.~\ref{Fig:cellAR}, we present data for several coefficients of friction, and for two values of the particle aspect ratio, $\alpha=1.5$ and $2$. For each case, we prepared simulations with four different cell aspect ratios. The symbols show the results of the DEM simulations, while the theoretical prediction is represented by the dashed lines. To present the response of the system, we use a rescaled value, namely $a_{y\text{,rescaled}}=a_y(L_y^2+L_z^2)/(L_y^2 L_z^2)$, as suggested by the prefactor in expressions (\ref{Eq:finaly}) and (\ref{Eq:finalz}). 
This way, the role of the cell aspect ratio is highlighted, and the effect of the actual dimensions of the cell is ruled out, facilitating the comparison between the different systems. One can see the good qualitative agreement between the simulated and calculated data in terms of changes in the vortex amplitude as a function of $L_z/L_y$.
In addition, the inversion can also be tracked in Appendix~\ref{App:th} in Figs.~\ref{Fig:theory} and \ref{Fig:theory2} where the quantities $\mu$, $\nu_2$, $\pi$ and $a_y$ are presented as a function of $\mu_p$ and $\alpha$.

Further investigating the data, we see that the $a_y$ versus $L_z/L_y$ curves are not independent of microscopic parameters. For example, in Fig.~\ref{Fig:cellAR}a, for the numerical data, the zero-crossing is observed at a larger cell aspect ratio for smaller $\mu_p$. This means that for an intermediate but fixed $L_z/L_y$, the sign of $a_y$ can be changed when varying the particle friction coefficient. The theoretical lines display similar behavior, but for a much more restricted range of $L_z/L_y$, and since the lines cross the other way round, the effect is opposite [see Appendix~\ref{App:th} Fig.~\ref{Fig:theory2}(a)]. Finally, note that each point obtained by DEM simulations is an average of five independent realizations, but the difference between these realizations is small, and we have not seen a change of sign between the different realizations corresponding to the same simulation parameters. For comparison, we present the $a_x$ versus $L_z/L_y$ curves in Fig.~\ref{Fig:cellAR_x} of Appendix~\ref{App:th}.

\begin{figure*}
\centering
\includegraphics[width=0.99\textwidth]{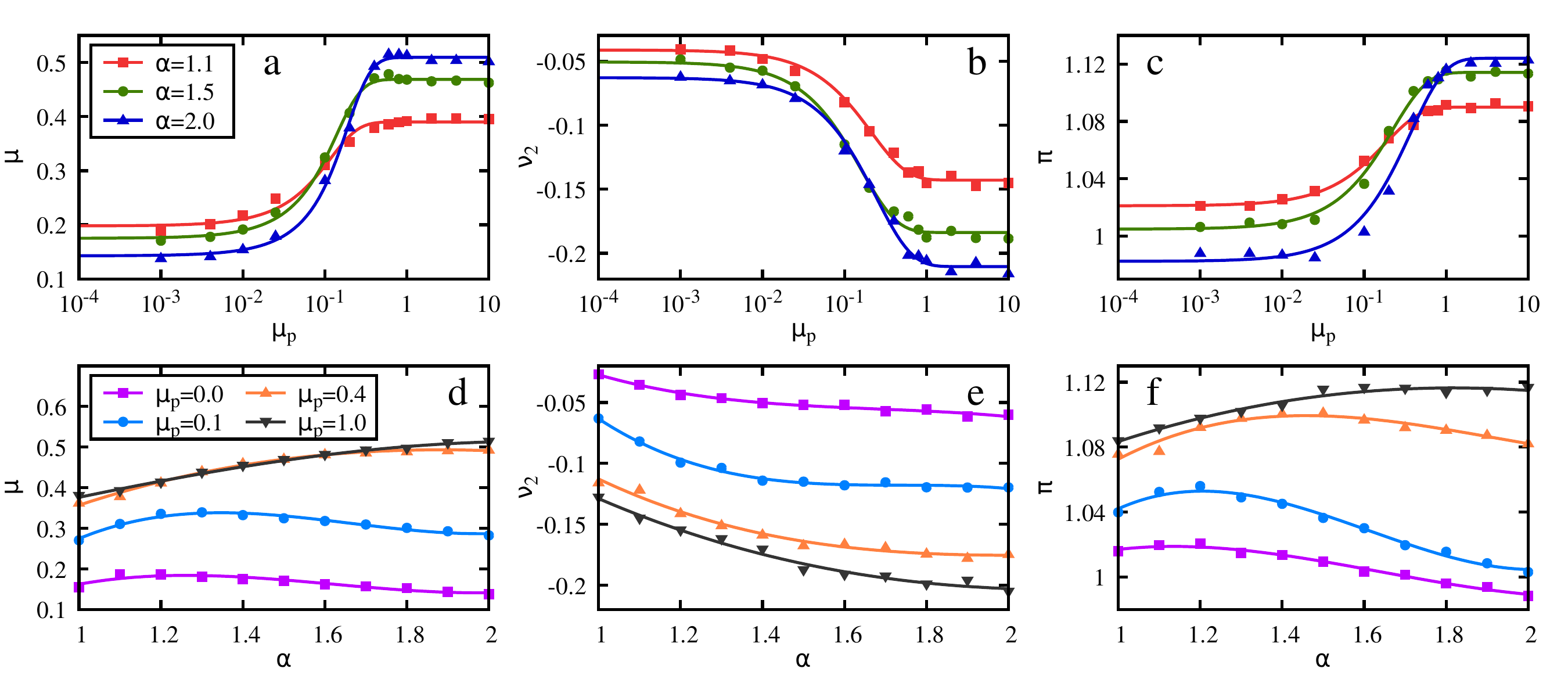}
\caption{(a), (d) $\mu$, (b), (e) $\nu_{2}$, and (c), (f) $\pi$ as a function of the particle-particle coefficient of friction (a)-(c) ($\mu_p$) and (d)-(f) the particle aspect ratio ($\alpha$) for cell aspect ratio $L_z/L_y=1$ and inertial number $I=10^{-2}$ using results of Nagy \textit{et al.}~\cite{nagy2020flow}.}
\label{Fig:theory}
\end{figure*}

We note that the agreement between simulations and theory is qualitative only: the response measured in the numerical simulations is typically of smaller amplitude than what the theory always predicts. This is why we used distinct vertical axes in the figures for the magnitude of the velocities. We also see that the difference between the simulation results and the theoretical prediction for the transverse direction is larger for large $L_z/L_y$ (order of 20) than for small $L_z/L_y$ (order of two to three), and that the inversion point, i.e. the zero-crossing of the curves in Fig.~\ref{Fig:cellAR}, is observed at a larger $L_z/L_y$ than predicted. As for the response in the shear direction (Fig.~\ref{Fig:cellAR_x}), the difference between the theory and numerics does not depend on the cell aspect ratio, it is in the order of 20. The lack of quantitative matching could be thought to be surprising, given the fact that we are using calibrated parameters $\mu_c$, $\nu_{1c}$ and $\nu_{2c}$ of the rheological law on very similar numerical runs (in the homogeneous case, i.e. with $F_x=0$)~\cite{nagy2020flow}. The possible adjustment of the remaining free parameter $\eta_c$ does not improve much the matching, as the theoretical results are in practice not very sensitive to its value, and all presented curves have been computed with $\eta_c$ (or $\beta_c$) equals $0$. We do not think that the rheological parameters of \cite{nagy2020flow} should be reassessed. In fact, several reasons can be invoked to explain this quantitative difference. First, as mentioned in the previous section, even for a perturbation amplitude as small as possible to measure a transverse velocity response on the order of a few percent, the corresponding response in the $x$ direction is large, on the order a several tens of percent (see Fig.~\ref{Fig:cellAR_x} in Appendix~\ref{App:th}), i.e. significantly above the linear regime. These nonlinearities are likely to be the main cause of the quantitative mismatch between simulations and theory. A second reason is that the theory is for continuous fields, while the simulations are performed with discrete grains. This leads to the question of whether or not the ratio between the system and the grain sizes is large enough. For the base homogeneous system, the transverse velocity gradient takes place over $\simeq 20$ grains, which seems to be reasonable. However, the heterogeneities we induce by the forcing $F_x$ operate at a scale typically four times smaller, i.e., over $\simeq 5$ grain diameters. Clearly, we are here at the limit of a continuum. Strong spatial gradients are also known to induce nonlocal effects \cite{PouliquenPTRSA2009, BouzidPRL2013, HenannPNAS2013}, which are not accounted for in the present theory. A good test would be, of course, to run simulations with a larger number of grains, especially in the case of more elongated particles. This is, however, computationally expensive, especially as averaging needs to be performed over rather long deformations. An extension of the present theory to account for the first nonlinear and/or nonlocal terms is a piece of work in itself, which we do not address here and leave for a future study. A last reason to explain the predicted versus measured quantitative difference is that the theory assumes incompressibility. However, as the packing fraction maps show [Fig.~\ref{Fig:cellAR_profiles}(b)], this is not the case in the DEM simulations. Clearly, lower packing fractions correspond to higher velocity gradients [Fig.~\ref{Fig:cellAR_profiles}(c)]. These heterogeneities are likely to result from the strong stream-wise perturbation and can have significant feedback on the secondary flows and vortex amplitude, in a nonobvious way that needs to be further investigated, but which is beyond the scope of this paper. Note finally that accounting for incompressible $\mu(I)$ rheology for granular flows requires advanced techniques such as those described in~\cite{BarkerPRSA2017, BarkerJFM2017}, leading to a substantial reformulation of the theory.

Despite this quantitative mismatch between theory and simulations, evidence for the inversion of the vortex pattern and its geometrical origin is a major result of the present paper. Also, following the rheological interpretation of the numerical measurements, we understand that these secondary flows are induced by the normal stress differences, and in particular by the second one, as the factor $\nu_{2c}$ is present in the expressions of the vortex amplitudes $a_{y}$ and $a_{z}$ [see Eqs. (\ref{Eq:finaly}) and (\ref{Eq:finalz})]. Interestingly, a buoyancy-driven explanation for secondary flows in Couette cells was proposed by \cite{krishnarajNC2016}. Here, although we do observe a packing fraction pattern associated with the vortices [Fig.~\ref{Fig:cellAR_profiles}(b)], it cannot generate convection as there is no gravity in the system. In addition, we do not observe any qualitative change in these packing fraction maps, nor for those of the velocity gradient [Fig.~\ref{Fig:cellAR_profiles}(c)], when the cell aspect ratio increases, associated with the vortex pattern inversion.

\begin{figure}[b]
\centering
\includegraphics[width=0.48\textwidth]{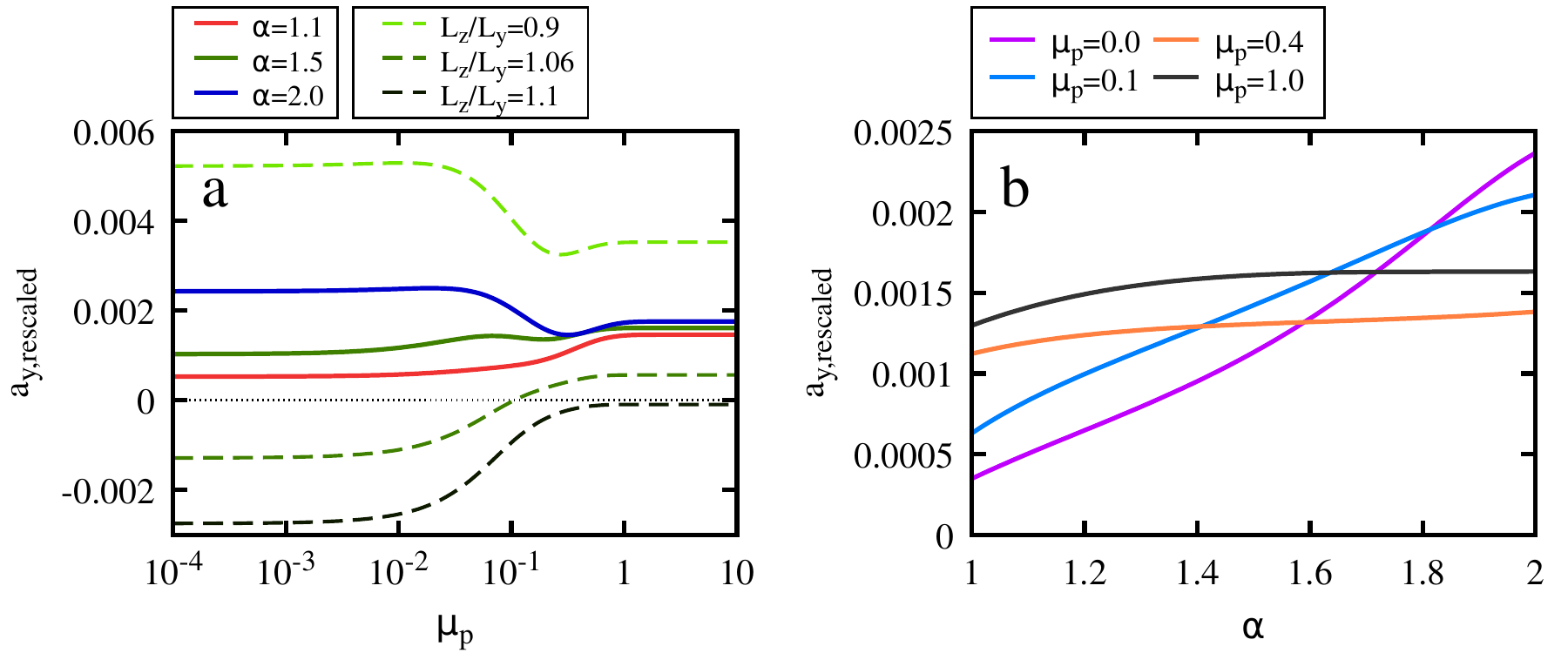}
\caption{The rescaled response of the system [$a_{y\text{,rescaled}}=a_y(L_y^2{+}L_z^2)/(L_y^2 L_z^2)$] as function of (a) the particle-particle coefficient of friction ($\mu_p$) and (b) the particle aspect ratio ($\alpha$) for cell aspect ratio $L_z/L_y=1$ (continuous lines), and $I=10^{-2}$. In panel (a) for $\alpha=1.5$, we show the influence of a slight change ($\pm10$\%) of the cell aspect ratio on the response of the system (dashed green lines). For cell aspect ratios slightly larger than $1$, we have a change of sign in the response.}
\label{Fig:theory2}
\end{figure}

\section{Conclusion and perspectives}
We studied the response of a granular shear flow of elongated particles to a perturbative streamwise forcing, running DEM numerical simulations and analytically computing the linear flow response from the granular rheology generalized to this type of grain. Both reveal the emergence of transverse secondary flows that, upon averaging over sufficiently long times, generate a four-vortex pattern when the perturbation is a single Fourier mode. A central result is the observation and understanding of the inversion of this pattern when the cell aspect ratio is varied. We also observe a pattern inversion when the particle contact friction is changed at a fixed cell aspect ratio, but for a limited range. 

We discussed the possible reasons why the agreement between simulations and theory is qualitative only. This calls for longer simulations with a smaller perturbation amplitude or for runs with larger systems to check finite-size effects. In particular, we expect that gradients that are too strong at the scale of a few grains would not match well with the continuum theoretical approach and/or could be associated with nonlocal effects, which the present theory does not account for. Such tests are, however, computationally expensive, as a long time average is necessary to see the emergence of these vortex patterns out of the noise. We leave more systematic studies with such additional simulations for later work.

Finally, this work offers additional insight into the heaping problem observed in split-bottom shear devices~\cite{wortelSM2015,fischerNJP2016}. This phenomenon, which also shows secondary vortex flows, is, of course, more complicated than what we observe here in the perturbed shear cell configuration. Not only is the split-bottom geometry less easy to handle, but also because gravity is present and can affect these flows through buoyancy convection associated with spatial variations of the packing fraction. Nevertheless, it is striking that the amplitude of this heaping is sensitive to the filling height, i.e., the equivalent of the cell aspect ratio here. Also, the understanding that varying a microscopic parameter, such as the contact friction coefficient $\mu_p$, can invert the vortex pattern under some conditions provides a possible interpretation of the observation that nonconvex granular particles generate dipping rather than heaping in the split-bottom cell \cite{mohammadiPRE2022}. Clearly, more numerical simulations as well as experiments with different grains and geometries to investigate the effect of varied normal stress differences would be very instructive.

\begin{figure}[t]
\centering
\includegraphics[width=0.5\textwidth]{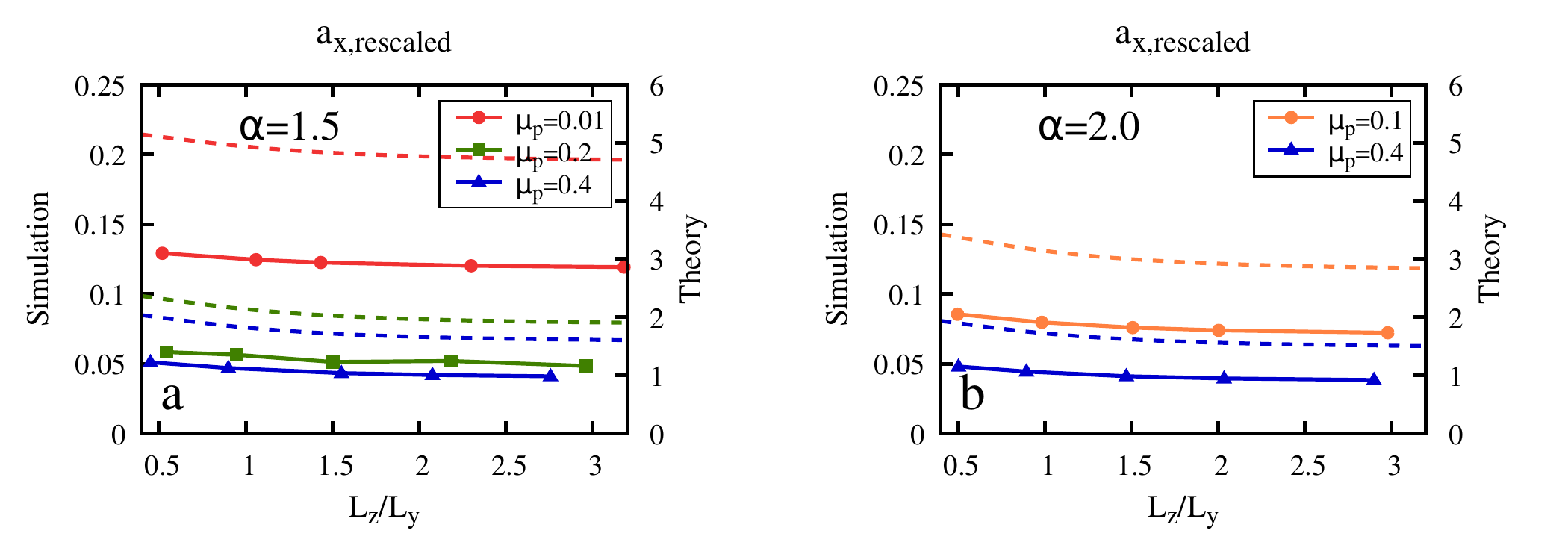}
\caption{Response of the system in the shear direction for the same simulations as presented in Fig.~\ref{Fig:cellAR}. The response is rescaled by the system size ($a_{x\text{,rescaled}}=a_x/L_z^2$), and plotted as a function of cell aspect ratio for two particle aspect ratios (a) $\alpha=1.5$ and (b) $\alpha=2$, and for different particle coefficients of friction $\mu_p$ (legends). The strength of perturbation is $A=2\times10^{-4}$. Dashed lines show the expected values according to the theory. The symbols show the results of simulations, averaged over five different realizations. Their span is in the order of the point size. Importantly, note the different vertical axes on these graphs for simulation results (left) and theoretical predictions (right) --- a qualitative-only agreement is reached.}
\label{Fig:cellAR_x}
\end{figure}

\section*{Acknowledgments}
The authors thank D\'aniel B\'alint Nagy for the input related to the generalized granular rheology. We also thank Pierre Jop, Cyril Gadal, and Julian Talbot for useful discussions. We acknowledge support from CNRS PICS/IEA Grant No. 08187, the Hungarian Academy of Sciences Grant No. NKM-2018-5, the COST action ”ON-DEM” (Grant No. CA22132), and the Spanish Government through Grant No. PID2023-146422NB-I00 supported by MICIU/AEI/10.13039/501100011033 and FEDER, EU.

\section*{Data availability}
The data that support the findings of this article are not publicly available upon publication because it is not technically feasible and/or the cost of preparing, depositing, and hosting the data would be prohibitive within the terms of this research project. The data are available from the authors upon reasonable request.

\appendix

\section{Rheological parameters}
\label{App:th}
Nagy \textit{et al.}~\cite{nagy2020flow} studied the rheological parameters, like the effective friction coefficient ($\mu$) and the normal stress differences ($N_{1}$ and $N_{2}$) of elongated grains in a plane Couette flow for various particle aspect ratios ($\alpha$), particle friction coefficients ($\mu_p$), and inertial numbers ($I$). In Fig.~\ref{Fig:theory} we present their measurements for the case of $I=10^{-2}$, and the interpolated values of the rheological parameters as functions of particle friction coefficient [Figs. \ref{Fig:theory}(a) to \ref{Fig:theory}(c)] and aspect ratio [Figs. \ref{Fig:theory}(d) to \ref{Fig:theory}(f)]. We remind the reader that $\nu_1 = \frac{N_1 + N_2}{\Pi}$, $\nu_2 = \frac{N_2}{\Pi}$, and $\pi = 1 - \tfrac{1}{3}(\nu_{1} + \nu_{2})$. Using these results, we calculated the linear response of our system ($a_{y\text{,rescaled}}$), which is presented in Fig.~\ref{Fig:theory2}.

\section{Time evolution of the secondary flow pattern}
To show the time evolution of the secondary flow vortex pattern, in the Supplemental Material (SM)~\cite{SM} we present the time evolution of the velocity fields for two cell aspect ratios presented in Fig.~\ref{Fig:cellAR_profiles}(a), for $L_z/L_y\approx0.5$ (SM Video 1,~\cite{SM}) and $L_z/L_y\approx1$ (SM Video 2,~\cite{SM}). The flow pattern with four vortices starts to form relatively quickly; however, the strength and relative position of the vortices change a lot as the system evolves.

\bibliographystyle{apsrev4-1}
\bibliography{refs}

\clearpage

\end{document}